\documentclass[aps,epsf,eqsecnum,feynmp,graphics,amsfonts,amssymb,latexsym,amsmath,twocolumn]{revtex4}
\usepackage[latin1]{inputenc}
\usepackage{graphics}
\usepackage{amsfonts, amssymb, latexsym, amsmath}
\input epsf.sty
\def\aut#1{#1}
\def\comment#1{}

\newcommand{\Bf}[1]{\mbox{\boldmath $#1$}}
\newcommand{\Bfs}[1]{\mbox{\scriptsize\boldmath $#1$}}
\begin{document}
\title{
End-To-End
Distribution Function Function
of Stiff Polymers\\ for all Persistence Lengths
}
\author{B.~Hamprecht and H.~Kleinert}
\address{Institut für Theoretische Physik, Freie Universität Berlin,\\
Arnimallee 14, D-14195 Berlin, Germany\\
{\scriptsize e-mails: bodo.hamprecht@physik.fu-berlin.de}
{\scriptsize e-mails: hagen.kleinert@physik.fu-berlin.de}}
\begin{abstract}
We set up
 recursion relations
for calculating
all
 even moments of the end-to-end distance of
a Porod-Kratky wormlike chains in $D$ dimensions.
From these  moments
we derive a simple analytic
expression
for
the end-to-end
distribution in three dimensions
valid for all peristence lengths.
It is
in excellent agreement with Monte Carlo
data for stiff chains and goes properly
over into the Gaussian random-walk distributions
for low stiffness.
\end{abstract}
\maketitle
\section{Introduction}
The main quantity of interest
for a stiff polymer  in $D$ dimensions
is the end-to-end distribution
defined by the path integral of a stiff polymer \cite{Hagen}
\begin{eqnarray} \!\!\!\!\!\!\!\!\!\!\!\!\!\!\!\!\!\!\!\!
P_L({\bf R})
&\!\!\!\propto\!\!\!&
\int d^Du_b
\int d^Du_a
\int {\cal D}^Du\,\nonumber \\&& \!\!\!\!\!\!\!\!\!\!\times
 \delta ^{(D)}
        \left( {\bf R}-\int_0^L ds\,
         {\bf u}(s)\right)\,
     e^{ -({\bar\kappa }/2)\int _0^Lds\,
[{\bf u}'(s)]^2}.
\label{15.82nc}
\label{@}\end{eqnarray}
where
 $\bar  \kappa $ is the reduced stiffness
related to the persistence length $\xi$ by
\begin{equation}
 \bar\kappa =\frac{ \kappa }{k_BT}=(D-1)\frac{\xi}{2},
\label{@batk}\end{equation}
The unit vectors ${\bf u}(s)$ are the tangent vectors of the
space curve of the polymer.
A Fourier representation of the $ \delta $-function brings this to the form
\begin{eqnarray}\!\!\!\! \!\!\!\!\!\!\!\!\! \!
P_L({\bf R})
&\!\!\!\!\propto\!\!\!\!&
\int _{-i\infty }^{i\infty }  \!
\frac{d ^D \lambda}{2\pi i}
 e^{
\bar  \kappa
       \Bfs{\lambda} {\bf R}/2
}\!\!\!
\int d^Du_b \! \!
\int d^Du_a
({\bf u}_bL|{\bf u}_a\,0)^{\Bfs \lambda }\!,
\label{15.82nc2}
\end{eqnarray}
where
\begin{eqnarray}
({\bf u}_b\,L|{\bf u}_a\,0)^{\Bfs \lambda }\!\!\equiv \!\!\!\int
_{{\bf u}(0)={\bf u}_a}
^{{\bf u}(L)={\bf u}_b}
\!{\cal D}^Du\,
     e^{ -({\bar\kappa }/2)\int _0^Lds\,\left\{
[{\bf u}'(s)]^2
+  \Bfs{\lambda} \cdot {\bf u}(s)\right\} }   \!\!\!\!
\nonumber \\&& \!\!\!\!\!\!\!
\label{15.82nc3}
\end{eqnarray}
coincides with the euclidean path integral
of a point particle of mass $M=\bar  \kappa $
moving on a unit sphere
in an external electric field ${\Bf \lambda }$.

Since ${\bf u}(s)$ are unit vectors,
the path integral is not solvable
exactly, except
for zero ${\Bf  \lambda }$.
It is however, easy to find
arbitrarily high even moments
of the end-to-end distance
of the distribution
$
P_L({\bf R})$:
\begin{eqnarray}
 \left\langle { R}^{2n}\right\rangle
 \equiv
 \int d^DR\,{ R}^{2n}
P_L({\bf R}).
\label{@mome}\end{eqnarray}
The $2n${th} moment  of the chain
can be obtained directly from the expansion coefficient
in powers of $\Bf  \lambda $
of
the integral over  (\ref{15.82nc3}).

In naturla units with
$       \bar \kappa)1$,
the path integral
(\ref{15.82nc3}) solves the
Schr\"odinger equation
\begin{eqnarray}
\left(-\frac{1}{2} \Delta _{\bf u}+ \frac{1}{2}\Bf  \lambda \cdot {\bf u}
+\frac{d}{d\tau}\right)
({\bf u}\,\tau|{\bf u}_a\,0)^{\Bfs \lambda }=0,
\label{@}\end{eqnarray}
where $ \Delta $ is the Laplacian on a unit sphere.
The
 external electric field
${\Bf  \lambda }$ may be assumed to point in the $z$-direction,
or the
 $D$th direction
in $D$-dimensions.
In the distribution (\ref{15.82nc2}), only  the integrated expression
\begin{eqnarray}
\psi(z,\tau; \lambda )\equiv \int d^Du_a({\bf u}\,\tau|{\bf u}_a\,0) ^{\Bfs \lambda }
 \label{@1.6}\end{eqnarray}
appears, which is a function
of $z=\cos \theta$ only, where $\theta$ is the angle
between ${\bf u}$ and the electric field ${\Bf \lambda }$.
Setting $\bar  \kappa =1$, for a moment,
we
obtain the
simple Schr\"odinger equation in euclidean time
\begin{eqnarray} \!\!\!\!\!\!
\hat H\psi(z,\tau; \lambda )=
-\frac{d}{d\tau}\psi(z,\tau; \lambda ),
\label{15.82nc4}
\end{eqnarray}
\begin{eqnarray} \!\!\!\!\!\!\!
\hat H&\equiv & \hat H_0+\lambda ~\hat H_I
=\!\!\!\!-\frac{1}{2}\Delta+\frac{1}{2}\lambda ~z
\nonumber \\
&=&-\frac{1}{2}\left[(1-z^2)\frac{d^2}{dz^2}-(D-1)z\frac{d}{dz}\right]+
\frac{1}{2}\lambda ~z
\end{eqnarray}

Now the desired
moments (\ref{@mome})
can be obtained
from the coefficient of $\lambda ^{2n} /2 ^{2n} (2n)!$
 in
the expansion of
the integral over  (\ref{@1.6}),
\begin{eqnarray}
f(\tau ;\lambda )\equiv \int_{-1}^1 dz \,
\psi(z,\tau ; \lambda ),
\label{@}\end{eqnarray}
in powers of $ \lambda $, evaluated at the euclidean time
$\tau = L$.

\section{Recursive Solution of the Schroedinger Equation.}
The function
$ f(L ; \lambda )$
has a spectral representation
\begin{eqnarray}
 f(L ; \lambda ) \!
\equiv \!\sum_{l=0}^\infty\frac{\int_{-1}^1dz\,\varphi^{(l)}{}^\dagger(z)
 \exp{\left(-{E^{(l)}L}\right)}
~ \varphi^{(l)}(0)}{\int_{-1}^1dz\,
\varphi^{(l)}
{}^\dagger(z)~\varphi^{(l)}(z)},
\label{II1}\end{eqnarray}
where $
\varphi^{(l)}(z)$ are the solutions of the time-independent
Schr\"odinger equation
$\hat H
\varphi^{(l)}
(z)=E^{(l)}
\varphi^{(l)}
(z)$.
Applying perturbation theory to this problem, we start from the eigenstates of the
unperturbed Hamiltonian $\hat H_0=-\Delta /2$, which are given by
the Gegenbauer polynomials $C_l^{D/2-1}(z)$
with the eigenvalues $E_0^{(l)}=l(l+D-2)/2$~\cite{Hagena}. Next we set up a recursion scheme for the perturbation expansion of the eigenvalues and eigenfunctions as described in \cite{HaPe}.
We begin with a brief review of the method.
Starting point is the usual expansion of energy eigenvalues
and states in powers of the coupling constant $\lambda$:
\begin{align}
\label{II2}
E^{(l)}
=&\sum_{j=0}^\infty ~\epsilon^{(l)}_j~\lambda ^j,~~~~
\\
\label{II3}
|\varphi^{(l)} \rangle =&\sum_{l',i=0}^\infty~\gamma^{(l)}_{l',i}~\lambda ^i~\alpha_{l'}\left|l'\right\rangle~.
\end{align}
The wave functions
$\varphi^{(l)}(z)$ are the scalar products
$\langle z|\varphi^{(l)} \rangle$.
The index $i$ counts the order
of the interaction strength $ \lambda $.
The lowest expansion coefficients
of the energy are of course
$ \epsilon _0^{(l)}=E_0^{(l)}$.
In the second line, we have introduced
auxiliary
normalization  constants $\alpha_{l'}$
 for convenience to be fixed later.
The state vectors  $|l\rangle$ of the
unperturbed
system are normalized to unity, but the state vectors
$|\varphi^{(l)}\rangle$
of the interacting system will be normalized in such a way, that
 $\langle\varphi^{(l)}|l\rangle=\alpha_l$
holds to all orders, implying that
\begin{equation}
\label{GAMMA}
\gamma_{l,i}^{(l)}=\delta_{i,0} \qquad \gamma_{k,0}^{(l)}=\delta_{l,k}\,.
\end{equation}
Inserting the above
expansions
into the Schrödinger equation, projecting the result onto the base vector $\langle k|\alpha_k$,
 and extracting the coefficient of $\lambda^i$, we obtain the relation:
\begin{equation}
\label{REC1}
\gamma_{k,i}^{(l)}\epsilon_0^{(k)}+
\sum_{j=0}^\infty \frac{\alpha_j}{\alpha_k}V_{k,j}\,\gamma_{j,i-1}^{(l)}=
\sum_{j=0}^i\epsilon_j^{(l)}\gamma_{k,i-j}^{(l)}\,,
\end{equation}
where $V_{k,j}= \lambda\langle k|z|j\rangle$
are the matrix elemts of the interaction
between unperturbed states.
For $i=0$,
 Eq.~(\ref{REC1}) is satisfied identically. For $i>0$,
 it leads to the following two recursion relations, one for $k=l$:
\begin{align}
\label{EQ1_1}
\epsilon_i^{(l)} =& \sum_{n=\pm1} \gamma_{l+n,i-1}^{(l)}W^{(l)}_{n} ,
\end{align}
 the other  for $k\ne l$:
\begin{align}
\label{EQ1_2}
\gamma_{k,i}^{(l)} =& \frac{\sum\limits_{j=1}^{i-1}
 \epsilon_j^{(l)}\gamma_{k,i-j}^{(l)}-\sum\limits_{n=\pm1} \gamma_{k+n,i-1}^{(l)}
W^{(l)}
_{n}}{\epsilon_0^{(k)}-\epsilon_0^{(l)}},
\end{align}
where only $n=-1$ and $n=1$ contribute to the sums over $n$ since
\begin{align}
W^{(l)}
_{n}\equiv &\frac{\alpha_{l+n}}{\alpha_l} \left\langle l\right|z \left|l+n \right\rangle=0,~{\rm for}~~n\neq\pm1.
\label{EQ1_3}
\end{align}
The vanishing
of  $
W^{(l)}
_{n}$ for $n\neq \pm1$ is due to
the
 band-diagonal form of the
matrix
of the  interaction $z$
 in the
 unperturbed basis $|l\rangle $. It is this property which
 makes the sums in (\ref{EQ1_1}) and (\ref{EQ1_2}) finite
and leads
to
recursion relations
with a finite number of terms
for all $\epsilon_i^{(l)}$ and $\gamma_{k,i}^{(l)}$.
To calculate
 $
W^{(l)}_{n}$, it is convenient
to express
$\left\langle l\right|z \left|l+n \right\rangle$
as matrix elements between  unnormalized
noninteracting
states $|l'\}$
 as
\begin{align}
\left\langle l\right|z \left|l+n \right\rangle=
\frac{\{ l|z|l+n \}}{\sqrt{
\{ l|l \}\{ l+n|l +n\}}}\,,
\label{EQ1_5}
\end{align}
where%
\begin{align} \!
\{ k|F(z)|l \}
\!\equiv \!\!\int_{-1}^1\!\!dzC_k^{D/2-1}\!(z)F(z)C_l^{D/2-1}\!(z)(1\!-\!z^2)^{(D\!-\!3)/2}\!,
\label{EQ1_6}
\end{align}
yielding
 \cite{Abramow}
\begin{eqnarray}
\{ l|l \}=\frac{2^{4-D} ~\Gamma(l+D-2)~\pi}{l!~(2l+D-2)~\Gamma(D/2-1)^2}
\label{EQ1_7}
\end{eqnarray}
Expanding the numerator of (\ref{EQ1_5}) with the help of the recursion relation for the
Gegenbauer polynomials \cite{RR}
\begin{align}
(l+1)|l+1 \}=(2l+D-2)\,z\,|l\}-(l+D-3)|l-1\}
\label{EQ1_8}
\end{align}
we find the only non-vanishing matrix elements to be
\begin{align}
\{ l+1|z|l \}=&\frac{l+1}{2l+D-2}\{ l+1|l+1 \},\\
\{ l-1|z|l \}=&\frac{l+D-3}{2l+D-2}\{ l-1|l-1 \}~.
\label{EQ1_8}
\end{align}
Inserting these together with (\ref{EQ1_7})
 into (\ref{EQ1_5}) gives
\begin{align}
\langle l|z|l-1 \rangle=
\sqrt{\frac{l(l+D-3)}{(2l+D-2)(2l+D-4)}},
\label{EQ1_a}
\end{align}
and a corresponding result for $\langle l|z|l+1 \rangle$.
We now fix the normalization constants
$ \alpha _{l'}$
by setting
\begin{align}
W^{(l)}
_{1}=\frac{\alpha_{l+1}}{\alpha_l}
\left\langle l\right|z \left|l+1 \right\rangle=1
\label{EQ1_b}
\end{align}
for all $l$,
which determines the ratios
\begin{align}
\frac{\alpha_{l}}{\alpha_{l+1}}= \left\langle l\right|z \left|l+1 \right\rangle=\sqrt{\frac{(l+1)~(l+D-2)}{(2l+D)~(2l+D-2)}}.
\label{EQ1_c}
\end{align}
Setting further
$\alpha_1=1$, we obtain
\begin{align}
\alpha_l=\left[ \prod_{j=1}^l~\frac{(2l+D-2)(2l+D-4)}{l(l+D-3)}\right]^{1/2}~.
\label{EQ1_9}
\end{align}
Using this we find
from (\ref{EQ1_3}) the remaining nonzero
$
W^{(l)}
_{n}$
for
 $n=-1$:
\begin{align}
W^{(l)}
_{-1}=&\frac{l(l+D-3)}{(2l+D-2)(2l+D-4)}\,.
\end{align}
We are now ready to  solve
the recursion relations
(\ref{EQ1_1}) and (\ref{EQ1_2}) for
$\gamma_{k,i}^{(l)}$ and $\epsilon_i^{(l)}$ order by order in $i$.
For the initial order $i=0$,
 the values of the $\gamma_{k,i}^{(l)}$ are given by Eq.~(\ref{GAMMA}).
The coefficients $\epsilon_i^{(l)}$ are equal
to the unperturbed energies $\epsilon_0^{(l)}=E_0^{(l)}=l(l+D-2)/2$. For
each $i=1,2,3,\dots\,$, there is only a finite number of non-vanishing
$\gamma_{k,j}^{(l)}$ and $\epsilon_j^{(l)}$
with  $j<i$ on the right-hand sides of
(\ref{EQ1_1}) and (\ref{EQ1_2})
which
allows us to calculate
$\gamma_{k,i}^{(l)}$ and $\epsilon_i^{(l)}$ on
 the left-hand sides.
In this way it is easy
to find the
perturbation expansions for  the energy
and the wave functions to high orders.\\

Inserting the resulting
expansions
(\ref{II2}) and (\ref{II3})
into Eq.~(\ref{II1}),
only the totally symmetric parts in $\varphi^{(l)}(z)$
will survive the integration in the numerators, i.e., we may
insert only
\begin{eqnarray}
\varphi^{(l)}_{\rm symm}(z)
=
\langle z|\varphi^{(l)}_{\rm symm}\rangle=\sum_{i=0}
^\infty\gamma^{(l)}_{0,i}\, \lambda ^i\left\langle z|0\right\rangle.
\end{eqnarray}
The denominators of (\ref{II1}) become explicitly
$
\sum_{l',i}|\gamma_{l',i}^{(l)}\,\alpha_{l'}|^2\,\lambda^{2i}$,
where the summation over $i$ is limited by
power of $\lambda^2$ up to which we want to carry the perturbation series;
also $l'$ is restricted to a finite number of terms only,
because of the band-diagonal structure of the $\gamma_{l',i}^{(l)}$.

Extracting the coefficients of the power expansion in $ \lambda $ from
(\ref{II1}) we
obtain all desired  moments
of the end-to-end distribution,
the
lowest two being, after reinserting $\bar\kappa $ from (\ref{@batk}),
\begin{equation}
\langle R^2\rangle = {2} \left\{
\xi L-{\xi ^2} \left[ 1-e^{-L/\xi }\right] \right\}.
\label{II5}
\end{equation}
\begin{eqnarray} \label{15.114}
\!\!\!\!\!&&\!\!\!\!\!\!\!\!\langle R^4\rangle
\!=\! \frac{4(D\!+\!2)}{D}L^2 \xi ^2
\!-  8L\xi ^3\left(
\frac{D^2\!+\!6D\!-\!1}{D^2}-\frac{D\!-\!7}{D\!+\!1}e^{-L/\xi}\right)
\nonumber \\  & & ~~
\!+4 \xi ^4
\left[
\frac{D^3+23D^2-7D+1}{D^3}
-2\frac{(D+5)^2}{(D+1)^2}e^{-L/\xi}
\right.\nonumber \\&&
\left.~~~~~~~~~~~~~~~~~~~~~~~~~~~~+2\frac{(D-5)^5}{D^3(D+1)^2}e^{-2DL/(D-1)\xi}
\right].       \nonumber
\end{eqnarray}%
The calculation of higher moments
is straightforward
with
a Mathematica
program, which we have made available
on the internet
in notebook form
\cite{NBK}.\\
\section{From Moments to End-to-End Distribution in $D=3$ Dimensions.}
\begin{figure}
\unitlength=1mm
\def\fsz{\footnotesize}
\def\ssz{\scriptsize}
\def\tsz{\tiny}
\def\pu#1#2{\put(#1,#2){\emmoveto}}
\def\pd#1#2{\put(#1,#2){\emlineto}}
\vspace{.5cm}
\hspace{-3.22cm}
\begin{picture}(117.86,59.99)
\def\IncludeEpsImg#1#2#3#4{\renewcommand{\epsfsize}[2]{#3##1}{\epsfbox{#4}}}
\def\IncludePCXImg#1#2#3#4{\unitlength#3mm\begin{picture}(#1,#2)\put(0,#2)
{\special{em:graph #4}}\end{picture}\vspace{0mm}}
\put(28.1,-9.1){\scalebox{.86}[.78]{\includegraphics{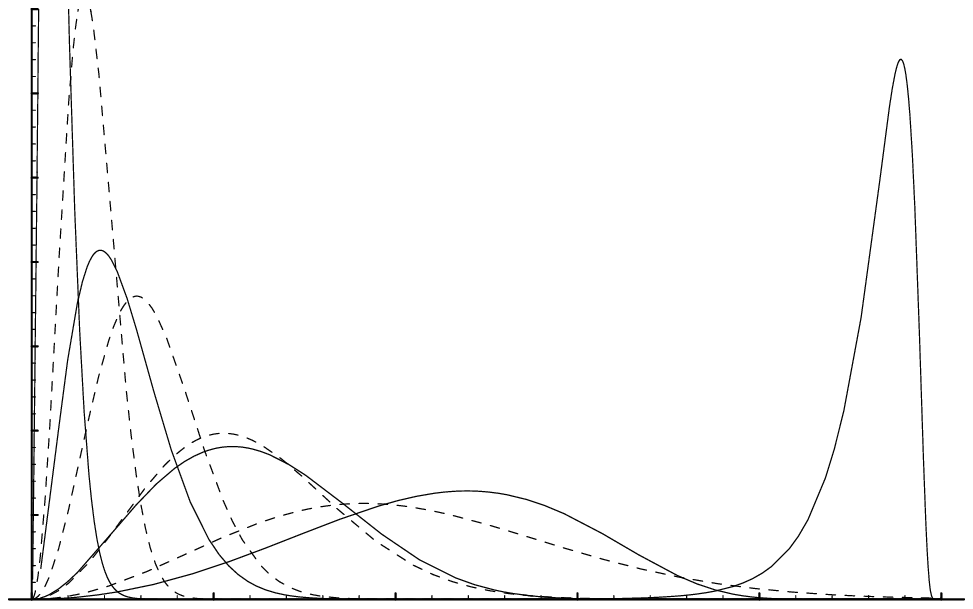}}}
\put(32.1,7.7){\scalebox{.82}[.786]{\includegraphics*{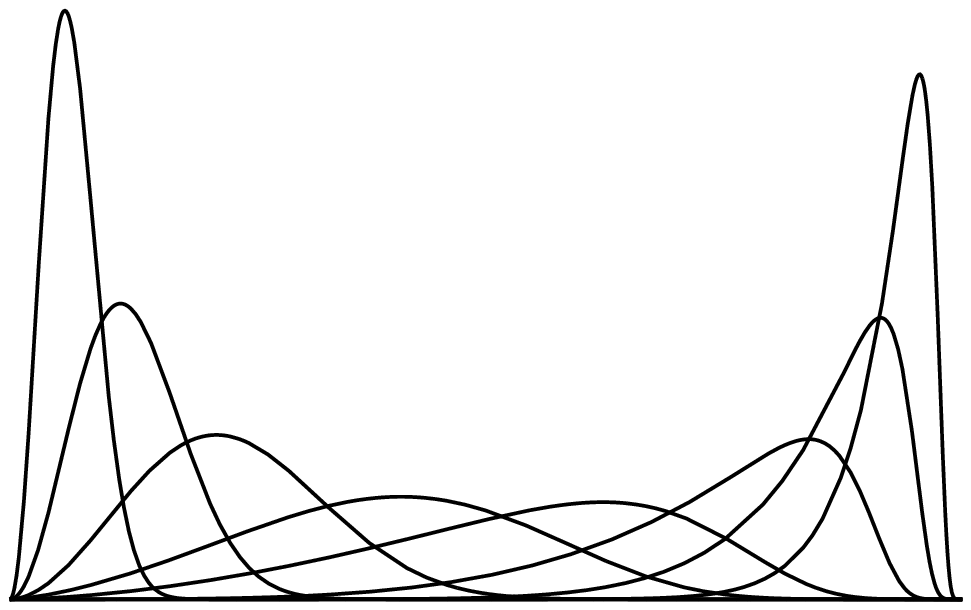}}}
\put(20,-0){\IncludeEpsImg{117.86}{62.99}{1.2000}{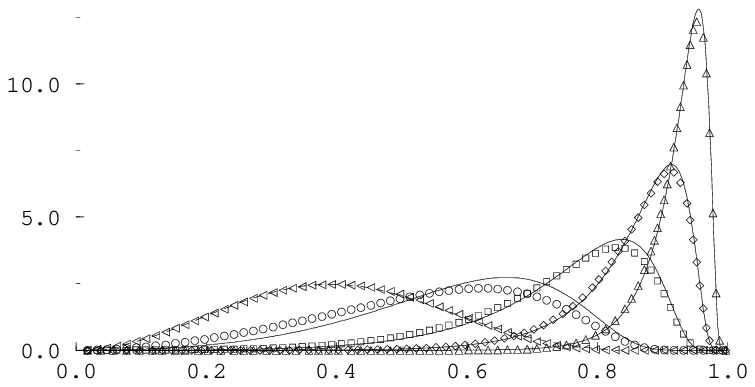}}
\put(69,3.5){\fsz $r=R/L$}
\put(64,45){\fsz $D=3,~~~~P_L({\bf R})$}

\put(42,39){\fsz 1/400}
\put(46.5,29){\fsz 1/100}
\put(52,22.8){\fsz 1/30}
\put(65,19){\fsz 1/10}
\put(82,19){\fsz 1/5}
\put(94.,22.8){\fsz 1/2}
\put(102,29){\fsz 1}
\put(105.3,39){\fsz 2}
\end{picture}
\vspace{-4mm}
\caption[F1]{Distribution of the end-to-end distances
of  polymer
 for different stiffnesses, parametrized
by the persistence lengths
$\xi=1/400,\,1/100,\,1/30,\,1/10,\,1/5,\,1/2,\,1,\,2$.
They are compared with
the Monte Carlo calculations
of Wilhelm and Frey \cite{Frey}
 (symbols) and with his
 large-stiffness one-loop perturbative results (thin curves).
For the small stiffnesses $\xi=1/400,\,1/100,\,1/30$, the curves
are well approximated by
Gaussian  random chain distributions
on a lattice with lattice constant
$a_{\rm eff}=2\xi$
which ensures that
$a_{\rm eff}=2\xi$
the lowest moments  $\left\langle R^2\right\rangle =
a_{\rm eff}L
$
are properly fitted (dashed curves).
}
\label{F1}
\end{figure}

The moments can  now be used to recover
the experimentally accessible
end-to-end distribution
of the polymers
 for various degrees of stiffness.
We parameterize
the distribution with an analytic form
\begin{align}
P_L({\bf R})\propto r^{k+2}(1-r^\beta)^m,~~~r\equiv R/L.
\label{@anal}\end{align}
whose  moments are
\begin{equation}
\displaystyle\left\langle { R}^{2n}\right\rangle=
\displaystyle\frac{\Gamma(\frac{3 + k + 2\,n}{\beta })\,
    \Gamma(\frac{3 + k }{\beta }+m+1)}{\Gamma(
     \frac{3 + k}{\beta })\,\Gamma(\frac{3 + k + 2\,n }{\beta }+m+1)}.
\end{equation}
We now adjust the  three parameters $k$, $\beta$, and $m$ to fit
the three most important moments of the distribution exactly,
 ignoring all others.
If the distances were distributed
uniformly over the interval $r\in [0,1]$,
the
moments would be $\left\langle R^{2n}\right\rangle ^{\rm flat}=1/(2n+2)$.
Comparing our exact moments $\left\langle R^{2n}\right\rangle (\xi)$
with the
flat ones, we find that $\left\langle R^{2n}\right\rangle (\xi)
/\left\langle R^{2n}\right\rangle^ {\rm flat}$ has a maximum for $n$
close to
$n_{\rm max}(\xi)\equiv 4\xi$.
The most important moments turn out to be
the ones
with
 $n=n_{\rm max}(\xi)$ and
$n=n_{\rm max}(\xi)\pm 1$.
If $n_{\rm max}(\xi)\le1$
we use
 the lowest even moments
$\left\langle R^2\right\rangle\!,
\,\left\langle R^4\right\rangle$, and $
\left\langle R^6\right\rangle$.
With these adjustments
the resulting distributions
are shown in Fig. \ref{F1}
for various
persistence lengths
 $\xi$.   They are in excellent agreement
with the Monte Carlo data (symbols)
obtained by Wilhelm and
Frey \cite{Frey}, and better than
their one-loop perturbative
results (thin curves)
which are good only for very stiff polymers.
For the small persistence
lengths $\xi=1/400,\,1/100,\,1/30$, the curves
are well approximated by
Gaussian  random chain distributions
on a lattice with lattice constant
$a_{\rm eff}=2\xi$
which ensures that
$a_{\rm eff}=2\xi$
the lowest moments  $\left\langle R^2\right\rangle =
a_{\rm eff}L
$
are properly fitted.

%
%

~~~\\
\vspace{2.7cm}
~\\ Acknowledgment\\
This work was partially supported by
ESF COSLAB Program and by the Deutsche Forschungsgemeinschaft
under Grant Kl-256.
\vspace{.5cm}

\end{document}